# Intrinsic Redshifts and the Tully-Fisher Distance Scale

David G. Russell

Owego Free Academy

Owego, NY USA

russeld1@oagw.stier.org

## Abstract

The Tully-Fisher relationship (TFR) has been shown to have a relatively small observed scatter of ~ +/-0.35 mag implying an intrinsic scatter < +/-0.30 mag. However, when the TFR is calibrated from distances derived from the Hubble relation for field galaxies scatter is consistently found to be +/-0.64 to +/-0.84 mag. This significantly larger scatter requires that intrinsic TFR scatter is actually much larger than +/-0.30 mag, that field galaxies have an intrinsic TFR scatter much larger than cluster spirals, or that field galaxies have a velocity dispersion relative to the Hubble flow in excess of 1000 km $s^{-1}$. Each of these potential explanations is contradicted by available data and the results of previous studies. An alternative explanation is that the measured redshifts of galaxies are composed of a cosmological redshift component predicted from the value of the Hubble Constant and a superimposed intrinsic redshift component previously identified in other studies. This intrinsic redshift component may exceed 5000 km $s^{-1}$ in individual galaxies. In this alternative scenario a possible value for the Hubble Constant is 55-60 km $s^{-1}$ $Mpc^{-1}$.

## Introduction

The Tully-Fisher relation (TFR - Tully&Fisher 1977) is the most important secondary distance indicator for determining distances to large numbers of spiral galaxies. There have been two conflicting approaches for using the TFR as a means of determining the value of the Hubble Constant. In one approach, it is believed that intrinsic TFR scatter is large (~ +/-0.70 mag) and therefore TFR distances are significantly affected by Malmquist bias (Bottinelli et al 1986, 1987,1988; Teerikorpi 1987; Federspiel et al 1994; Sandage 1994, 1999; Theureau et al 1997; Ekholm et al 1999). Supporters of this approach advocate large bias corrections and derive Hubble Constant values of 52 to 55 km $s^{-1}$ $Mpc^{-1}$ from the TFR (Ekholm et al 1999; Theureau et al 1997; Sandage 1999). In the second approach, it is believed that TFR scatter is small (~0.35 mag) and therefore bias corrections are small (Giovanelli et al 1997a,b; Tully&Pierce 2000; Sakai et al 2000). Supporters of this view find the Hubble Constant to be from 69 to 77 km $s^{-1}$ $Mpc^{-1}$ (Giovanelli et al 1997b; Tully&Pierce 2000; Sakai et al 2000).

Proponents of both positions present evidence and analyses that support their conclusions. Since the two points of view lead to contradictory results, both approaches cannot be correct. The resolution to this problem can be found by considering the different methods used to calibrate the TFR. In this paper it is shown that both points of view may be partially incorrect because the effects of non-velocity (intrinsic) redshifts are unaccounted for in the analysis methods.

While the subject of non-velocity redshifts has a controversial history, the purpose of this paper is not to argue in favor of or against any particular theoretical structure. Arp (see 1987, 1998 for reviews) has argued that the existence of intrinsic redshifts imply a non-expanding universe. Bell (2002,2004) has proposed that intrinsic redshifts are superimposed upon expansion of the universe. Lopez-Corredoira (2003) has offered a nice review of the matter. The subject and conclusions of this paper are compatible with both expanding and non-expanding universe models.

The evidence for intrinsic redshifts in spiral galaxies is discussed by Arp (1980, 1982, 1987,1988, 1990,1994,1998), Moles et al (1994), and Russell (2005). The evidence indicates that late type (Sbc and Sc) spiral galaxies tend to have excess redshift relative to early type (Sa/Sb) spirals in clusters and relative to the Hubble flow with the largest excess redshifts found for ScI type galaxies(eg Arp 1988,1990;Russell 2005).

The use of the TFR as a test for the existence of intrinsic redshifts is helpful because the distance determination is independent of the Hubble relation. Since ScI galaxies have been found to have the largest excess redshifts (Arp 1988, 1990), Russell (2005) determined the TFR distances to a sample of ScI galaxies from Mathewson&Ford (1996) and Giovanelli et al (1997a,b). Figure 1 is a plot of redshift (Vcmb) vs. TFR distance for the 102 ScI galaxies selected for the Russell (2005) analysis. The solid line represents a Hubble Constant of 72 km $s^{-1}$ $Mpc^{-1}$. The ScI galaxies have a strong tendancy to have larger redshifts than predicted by the Hubble Relation. It is important to note that the TF distances were calculated with the type dependent TFR (TD-TFR – see Russell 2004,2005). In traditional TFR calibrations ScI galaxies have systematically underestimated TFR distances (Russell 2004,2005) and therefore the excess redshifts of ScI galaxies seen in Fig. 1 would be even more extreme.

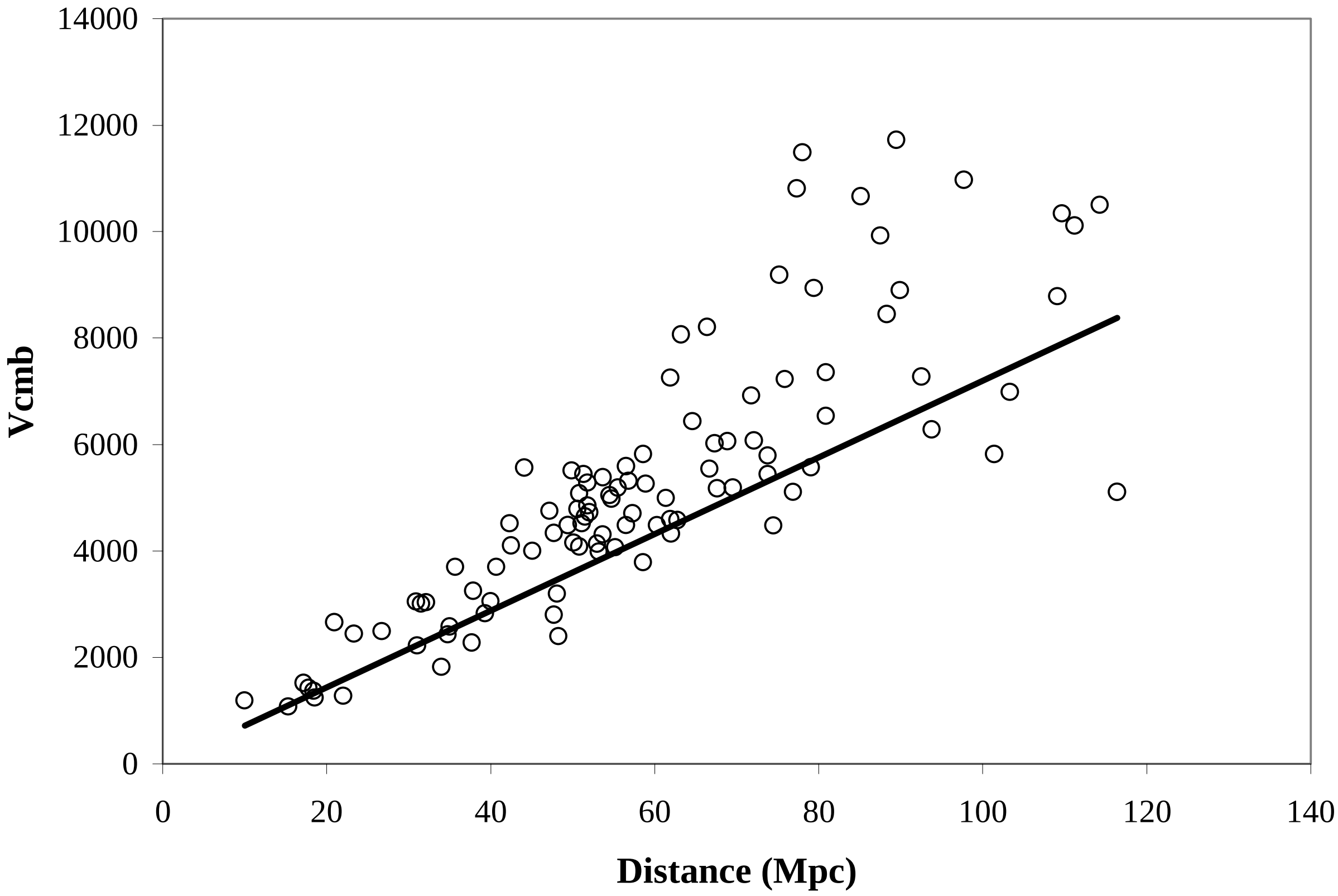

Figure 1 – Hubble Relation for 102 ScI group galaxies with TD-TFR distances (Russell 2005). Solid line represents a Hubble Constant of 72 km $s^{-1}$ $Mpc^{-1}$.

Table I is a list of the nine ScI galaxies from Russell (2005) that deviate most strongly from the Hubble relation if at their TF distances. Most of these galaxies have a redshift distance modulus that exceeds the TF distance modulus by more than 1.15 mag or an excess redshift from 3400 km $s^{-1}$ to 5900 km $s^{-1}$ if at the TFR distance. For example, ESO 445-27 has a TD-TFR distance of 89.5 Mpc and a redshift distance ($H_0$=72 km $s^{-1}$ $Mpc^{-1}$) of 163 Mpc. This discrepancy amounts to 1.30 mag in distance modulus or an excess redshift of +5278 km $s^{-1}$. Uncertainty in the B-band magnitude and rotational velocity result in an uncertainty of +/-0.294 mag in the ESO 445-27 distance modulus – which is too small to account for the 1.30 mag discrepancy between the redshift and TD-TFR distance moduli. Since excess redshifts over 3400 km $s^{-1}$ are too large to be real peculiar motions, Russell (2005) argued these discrepancies are caused by non-velocity intrinsic redshifts (See Russell 2005, section 7 for discussion on the problems with "mainstream" explanations). Additional evidence supporting an intrinsic redshift interpretation is discussed in Russell (2005b).

Table I: ScI's with extreme redshift anomalies

| Galaxy | logVrot | Incl. | m-$M_{TF}$ | $Mpc_{TF}$ | Vcmb | m-$M_{72}$ | $Mpc_{72}$ | $PV_{72}$ |
|---|---|---|---|---|---|---|---|---|
| 254-22 | 2.377 | 57 | 34.44 | 77.3 | 10813 | 35.88 | 150.2 | +5247 |
| 445-27 | 2.389 | 60 | 34.76 | 89.5 | 11722 | 36.06 | 162.8 | +5278 |
| 501-75 | 2.250 | 61 | 33.22 | 44.1 | 5563 | 34.44 | 77.3 | +2388 |
| 384-9 | 2.260 | 59 | 34.46 | 78.0 | 11487 | 36.01 | 159.5 | +5871 |
| 147-5 | 2.274 | 34 | 34.65 | 85.1 | 10666 | 35.85 | 148.1 | +4539 |
| 297-36 | 2.274 | 56 | 34.95 | 97.7 | 10972 | 35.91 | 152.4 | +3938 |
| 30-14 | 2.228 | 31 | 34.11 | 66.4 | 8207 | 35.28 | 114.0 | +3426 |
| 52-20 | 2.196 | 63 | 34.00 | 63.2 | 8068 | 35.25 | 112.1 | +3518 |
| 545-13 | 2.350 | 34 | 34.71 | 87.5 | 9923 | 35.70 | 137.8 | +3623 |

The galaxies in Table I are not simply anomalous exceptions with larger Tully-Fisher errors (Russell 2005, 2005b). They are the most extreme examples of the trend identified in Fig. 1 and were selected for illustration because their redshift discrepancies are clearly too large to be accounted for by peculiar motions. In addition, the required TFR errors are much larger than TF data uncertainties allow (Russell 2004,2005,2005b). *The important point is that all of the galaxies in Fig. 1 may contain a component of intrinsic redshift ranging from only a few km $s^{-1}$ to over 5000 km $s^{-1}$*. If real, then these intrinsic redshifts would have a significant impact on calculations of the Hubble Constant and would introduce large errors in distances calculated assuming a smooth Hubble flow. This is the topic of the following sections of this paper. Section 2 discusses observed Tully-Fisher scatter from the literature. Section 3 discusses possible explanations for the TFR scatter derived when calibrating with Hubble distances. Section 4 discusses implications for the value of the Hubble constant.

## 2. Observed Tully-Fisher Scatter

As noted in the introduction, there is a difference of opinion as to the correct value of the scatter in the TFR with reported scatter ranging from <+/-0.30 mag to >+/-0.80 mag. Table II summarizes the TFR scatter derived from major TFR studies. In an excellent review of the extragalactic distance scale Willick (1996) summarized the situation this way:

*"How can one reconcile this wide range of values? At least part of the answer lies in different workers preconceptions and preferences. Those excited at the possibility of finding a more accurate way of estimating distances tend to find low (σTF <= 0.30 mag) values. Those who doubt the credibility of TF distances tend to find high (σTF >= 0.50 mag) ones."*

But the key to the issue is the method of calibrating the TFR. Table II shows that the TFR scatter is much larger when Hubble distances are used to calibrate the TFR than when cluster samples and Cepheid calibrators are used. Since the effect of Malmquist bias is much greater when intrinsic TF scatter is greater, it is very important to determine which TFR calibration method is the best.

*Table II: Observed Tully-Fisher Scatter*

| Study | Observed Scatter | Calibration Method | Band |
|---|---|---|---|
| Tully&Pierce (2000) | +/-0.30 | Cepheid calibrators | B |
| | +/-0.23 | Cepheid calibrators | I |
| | +/- 0.38 | Cluster template | B |
| | +/- 0.34 | Cluster template | I |
| Giovanelli et al (1997) | +/-0.35 | Cluster template | I |
| Sakai et al (2000) | +/-0.43 | Cepheid calibrators | B |
| | +/-0.36 | Cepheid calibrators | I |
| Russell (2005) | +/-0.30 | Cepheid calibrators | B |
| | +/-0.09 to 0.16 | Cepheid calibrators TD-TFR | B |
| Russell (2004) | +/- 0.22 | Cluster scatter TD-TFR | B |
| | | | |
| Sandage (1994) | +/- 0.64 | Hubble distances ($H_0$=50) | B |
| Federspiel et al (1994) | +/- 0.59 | Hubble distances ($H_0$=50) | I |
| Persic&Salucci (1990) | +/-0.70 | Hubble distances ($H_0$=50) | B |
| Rubin et al (1985) | +/-0.68 | Hubble distances ($H_0$=50) | B |
| Corradi&Capaccioli (1990) | +/-0.84 | Hubble distances ($H_0$=75) | B |
| Kannappan et al (2002) | +/-0.73 to 0.82 | Hubble distances ($H_0$=75) | B |

The studies that find a small TF scatter use the Cepheid calibrators and cluster samples to calibrate the TFR (Giovanelli et al 1997a,b; Tully&Pierce 2000; Sakai et al 2000). For example, Tully&Pierce (2000 – hereafter TP00) used 155 galaxies in 5 clusters to create a template relation and then used 24 Cepheid calibrators to determine the zero point. The scatter derived from these methods ranges from +/-0.34 to +/-0.38 mag. The TFR can also be calibrated from the Cepheid calibrating sample alone (Sakai et al 2000, TP00) which results in scatter ranging from +/-0.30 mag to +/-0.43 mag.

Russell(2004,2005) split 27 TFR calibrators into two morphological groups. The ScI group includes galaxies of morphological types SbcI,SbcI-II,ScI,ScI-II,ScII and Seyfert galaxies. The Sb/ScIII group includes Sa to Scd galaxies not in the ScI group. It was demonstrated by Russell (2004,2005) that ScI group galaxies are systematically more luminous at a given rotational velocity than Sb/ScIII group galaxies. With a traditional single fit to all galaxies ScI group galaxies have systematically underestimated TF distances and Sb/ScIII group galaxies have systematically overestimated TF distances (Russell 2004,2005). To correct for this effect, the 27 calibrators were split into 12 ScI group galaxies and 15 Sb/ScIII group galaxies. The observed scatter was only +/- 0.09 mag for the ScI group, +/- 0.16 mag for the Sb/ScIII group, and +/- 0.22 mag for the 152 cluster galaxies in the cluster sample of Russell (2004).

The TFR has also been calibrated with Hubble distances for field galaxies where:

$$\text{Distance (Mpc)} = V_{cmb}/H_0 \qquad (1)$$

Absolute magnitudes are then determined from the redshift distance modulus and plotted against log Vrot to determine the TFR slope and zero point (eg Sandage 1994, Kannappan et al 2002). Sandage (1994) reported a scatter of +/- 0.64 mag for 249 field galaxies. Note that the scatter reported from the $H_0$ calibrated TFR is significantly larger (+/-0.59 to +/-0.84 mag – see Table II) than the scatter found from calibrators and cluster templates.

Figure 2 is a plot of the TFR calibration for the 102 ScI galaxies from Russell (2005) using Hubble distances to determine absolute magnitudes ($H_0$=72 km $s^{-1}$ $Mpc^{-1}$). As demonstrated in Table II, the ScI sample TFR scatter is significantly larger when using Hubble distances. The scatter for the 102 ScI galaxies (open circles in Fig 2) is +/- 0.46 mag compared with scatter of +/- 0.09 mag for the ScI calibrators (filled triangles in Fig. 2) and +/- 0.22 mag for the galaxies in the cluster sample of Russell (2004). Note that the scatter of +/-0.46 mag found here is smaller than the typical scatter from Hubble distances (~+/-0.70 mag) because the sample includes only ScI group galaxies. Including galaxies of all morphological types would increase the scatter because of the type effect.

Fig. 2 also demonstrates that if the Hubble distances are correct, the 12 ScI group calibrator galaxies from Russell (2004,2005) are less luminous at a given rotational velocity than the mean relation of 102 ScI galaxies. One interpretation of this offset is that the Hubble constant value used to calculate absolute magnitudes was too small. A Hubble constant of ~85 km $s^{-1}$ $Mpc^{-1}$ would eliminate the offset. However, the calibrators and the 102 ScI group galaxies have nearly identical TF slopes of 4.80 (dashed line in Fig. 2) and 4.73 (solid line in Fig. 2) respectively. Therefore, Hubble distances might be used to find the TF slope, but the calibrators must be used to find the zero point. In fact the zero point from the Hubble distances would be 20.87 compared with 20.43 for the calibrators.

In summary, observed scatter depends upon the calibration method. When using cepheid calibrators and cluster templates scatter is ~ +/-0.35 mag. When using the TD-TFR observed scatter is less than +/-0.25 mag. However, when using Hubble distances for field galaxies, the scatter is typically ~ +/-0.70 mag.

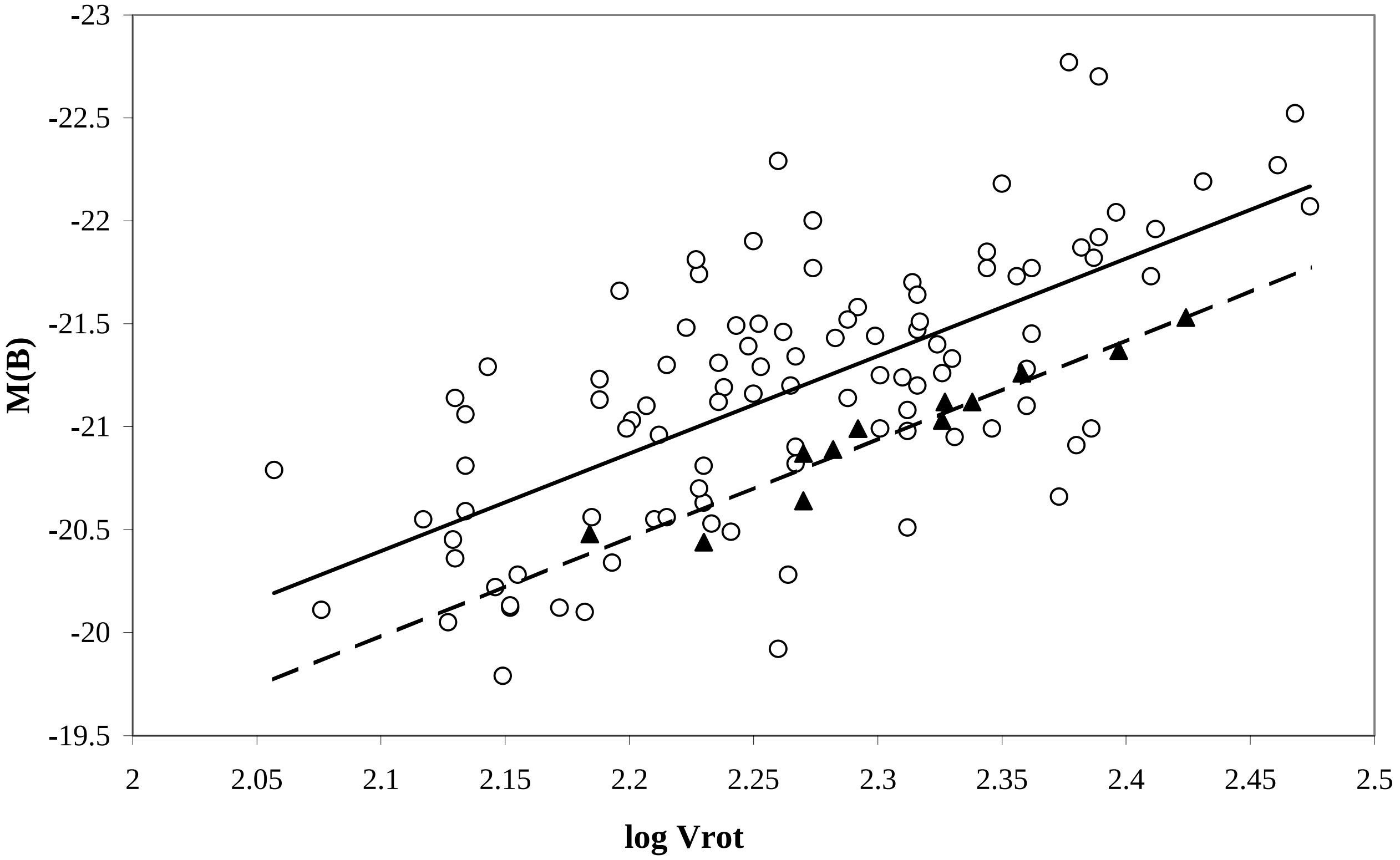

Figure 2 – Calibration of TFR for ScI galaxies using Hubble Distances ($H_0$=72 km $s^{-1}$ $Mpc^{-1}$). Open circles are ScI galaxies and solid line is least squares fit. Filled triangles are the 12 ScI group calibrators and dashed line is least squares fit. Note that (1) the calibrators would be systematically underluminous if the Hubble distances are correct and (2) the scatter with Hubble distances is significantly larger than found from calibrators and cluster samples.

# 3. Explaining the TFR scatter derived from Hubble distances

## 3.1 Is there a conventional explanation?

The interesting problem identified in section 2 is that the observed TFR scatter is significantly greater when using Hubble distances than when using cluster samples or the Cepheid calibrators. The larger TFR scatter found from Hubble distances has been used as justification for large bias corrections (Sandage 1994, 1999; Federspiel et al 1994, Theureau et al 1997, Ekholm et al 1999). When using Hubble distances to calibrate the TFR, peculiar motions introduce additional scatter. Both Sandage (1994) and Kannappan et al (2002) adopt a velocity dispersion from peculiar motions of ~200 km $s^{-1}$ for the field galaxies – a dispersion which Sandage describes as a "*gross overestimate*". At a distance of ~30 Mpc a 200 km $s^{-1}$ peculiar motion introduces an additional scatter of +/-0.25 to +/-0.30 mag to the distance modulus uncertainty. At 60 Mpc the introduced scatter would be only ~+/-0.15 mag.

The observed TF scatter from Hubble distances can be represented with the following equation:

$$\sigma_{obs} = (\sigma^2_{TF} + \sigma^2_{PV})^{1/2} \qquad (2)$$

where $\sigma_{obs}$ is the scatter observed when calibrating the TFR from Hubble distances, $\sigma_{TF}$ is the contribution to the observed TFR scatter from data errors and intrinsic TF scatter, and $\sigma_{PV}$ is the scatter contribution from peculiar motions. Note that $\sigma_{TF}$ should be equivalent to the typical scatter observed when calibrating the TFR from cluster samples and Cepheid calibrators (~+/-0.35 mag). Thus the predicted scatter observed when calibrating the TFR with Hubble distances should be equivalent to that expected if $\sigma_{TF}$ = ~+/-0.35 mag and $\sigma_{PV}$ = ~+/-0.25 mag. There might be some concern that different data sources may have different contributions to scatter from measurement errors, but there is no evidence in Table II of significant differences in TF scatter for the various studies that use Cepheid calibrators and cluster templates.

Sandage (1994) found $\sigma_{obs}$ = +/-0.64 mag and thus from equation 2 if $\sigma_{PV}$ = +/-0.25 mag., then $\sigma_{TF}$ = +/-0.59 mag. Kannappan et al (2002) found $\sigma_{obs}$ = +/-0.78 mag and therefore $\sigma_{TF}$ = +/-0.74 mag. After accounting for the contribution from peculiar motions, the resulting value of $\sigma_{TF}$ is still significantly larger than found from cluster templates and Cepheid calibrators. It is therefore necessary to re-examine the underlying assumptions of Sandage (1994). According to Sandage, the velocity dispersion of field galaxies relative to a smooth Hubble flow is at most 200 km $s^{-1}$ which implies that most galaxies are expected to follow a tight Hubble relation where Hubble distances agree within +/- 4 Mpc of the true distance. Sandage concludes that the large value of $\sigma_{obs}$ implies a large true intrinsic TF scatter:

*"Farther out in the extended region to v~2500 km $s^{-1}$, the dispersion in the random velocities is smaller than can presently be measured because all extant determinations depend on the differences between the kinematic and TF distances. Most of these differences must be interpreted not as real velocity perturbations but rather as errors in the photometric TF distances caused by the intrinsic dispersion in M(LW) of the TF relation, leading to systematic bias errors that increase with redshift –"*

Thus the key assumption underlying studies that calibrate the TFR from Hubble distances is that discrepancies between Hubble and TF distances are primarily due to intrinsic dispersion of the TFR. But is this assumption valid? Calibration of the TFR from Cepheid calibrators and cluster samples give a typical observed scatter of ~+/-0.25 to +/-0.40 mag - which demonstrates that this assumption is not valid. For the TD-TFR observed scatter is less than +/-0.25 mag. Sakai et al (2000) found that the chances of observing a scatter of only +/-0.25 mag if the true scatter is +/-0.70 is less than 1%.

The problem can be presented another way. If the observed TF scatter from Cepheid and cluster samples is ~+/-0.35 mag, then using field galaxies with Hubble distances should only add scatter from peculiar motions – at most +/-0.25 mag (Sandage 1994). Using equation 2 with $\sigma_{TF}$ = +/-0.35 mag and $\sigma_{PV}$ = +/-0.25 mag the observed TF scatter should only be +/-0.43 mag or less when the TFR is calibrated from Hubble distances. Why, then, is the scatter ~+/-0.70 mag when Hubble distances are used?

One possible explanation is that field spirals have a significantly larger intrinsic TFR scatter than cluster spirals. This explanation seems unlikely as TP00 noted that the five clusters utilized to create the cluster templates – Coma, Ursa Major, Pisces, A1367, and Fornax - represent very different environments and yet fit to the same TF plane. The Cepheid calibrators also come from a range of environments. There is no apparent reason why field galaxies should have significantly larger intrinsic TFR scatter.

An alternative possibility is that field spirals actually have much larger peculiar motions than presumed by Sandage (1994) and Kannappan et al (2002). Using equation 2, in order to account for $\sigma_{obs}$ of +/-0.64 mag from Hubble distances, if $\sigma_{TF}$ = +/-0.35 mag as found from the calibrators and cluster samples, $\sigma_{PV}$ equals +/-0.54 mag. which would require a velocity dispersion in excess of 1000 km $s^{-1}$. For the 102 ScI's in Fig. 2 +/- 0.21 mag is adopted as the upper limit of $\sigma_{TF}$ for the TD-TFR. With $\sigma_{obs}$ = +/-0.46 mag the resulting value of $\sigma_{PV}$ = +/- 0.41 mag which would require a velocity dispersion of ~ 900 km $s^{-1}$. There is no precedent for velocity dispersions this large in field galaxies with estimates for the local velocity dispersion ranging from only 25 km $s^{-1}$ to 75 km $s^{-1}$ (Karachentsev & Makarov 1996,2001; Ekholm et al 2001). Federspiel et al (1994) noted that peculiar motions cannot be responsible for the large observed TFR scatter found with Hubble distances because the redshift vs. distance plot is a wedge pattern in which scatter increases with increasing distance (Fig. 1 and see also Fig. 13 of Federspiel et al 1994). If peculiar motions were responsible for the TF scatter, then scatter should decrease with increasing distance. Federspiel et al claim that the wedge pattern results from intrinsic luminosity spread, but that explanation is at odds with the small scatter observed with calibration methods not involving Hubble distances.

The situation may be summarized this way: those who calibrate the TFR using Hubble distances find a large observed TFR scatter which they attribute to large *intrinsic* TFR scatter – an explanation contradicted by the small observed scatter found when using the calibrators and cluster samples. Those who calibrate the TFR from calibrators and cluster samples find a small observed TFR scatter which leads to a prediction that even with Hubble distances for field galaxies TFR scatter should be less than ~+/-0.45 mag if peculiar motions are at most 200 km $s^{-1}$. This expectation is contradicted by the large observed scatter (~+/-0.70 mag – but see Kannappan et al 2002 for even larger values) found when Hubble distances are used to calibrate the TFR. This discrepancy would require that the velocity dispersion of field galaxies be greater than 1000 km $s^{-1}$. A velocity dispersion this large has no precedent and is contradicted as a suitable explanation by the wedge shaped redshift–distance plots (Federspiel et al 1994). Thus the discrepancy between the two primary methods of calibrating the TFR may not be due to the methods used, but to something else entirely. The actual severity of this problem seems not to have been previously recognized in the literature.

### 3.2 Intrinsic Redshifts and TFR scatter

While the TF scatter derived from Hubble distances is problematic in conventional views, *the fact that the large TFR scatter arises only when Hubble distances are used implies that the source of the problem involves the observed redshifts*. The problem identified here is exactly what is expected if the observed redshifts of galaxies can contain a significant component of intrinsic redshift (Arp 1993, 1998; Bell 2002,2004). In these models the measured redshift results from the superposition of two primary contributions:

1. *Cosmological redshift component* – this component of the redshift is the traditional concept of a redshift-distance relation such that as distance increases redshift increases by an amount defined by some value of the Hubble Constant. The cosmological component is present whether the preferred model is an expanding universe model or not.
2. *Intrinsic redshift component* – this component of the redshift is non-cosmological and non-doppler in nature. In normal galaxies the intrinsic component may range from a few km $s^{-1}$ to thousands of km $s^{-1}$ and has a strong tendency to add to the cosmological component rather than subtract from the cosmological component. Whatever the mechanism behind intrinsic redshifts, the size of the intrinsic redshift appears to be generally correlated with the evolution or age of the galaxy such that younger galaxies tend to have larger intrinsic redshifts than older galaxies (Arp 1998; Bell 2002; Russell 2005).

It is important to note that in intrinsic redshift models there is still an underlying redshift-distance relationship from the cosmological component. For example, it was noted in section 1 that ESO 445-27 has a TD-TFR distance of 89.5 Mpc and a redshift of 11,722 km $s^{-1}$. If $H_0$=72 km $s^{-1}$ $Mpc^{-1}$, then ESO 445-27 has an excess redshift of +5278 km $s^{-1}$ or an excess of +6352 km $s^{-1}$ if $H_0$= 60 km $s^{-1}$ $Mpc^{-1}$. This excess is interpreted to be a non-velocity (intrinsic) redshift component and it follows that whatever value of the Hubble constant is adopted, the redshift distance to ESO 445-27 will be too large if it is assumed that all of the observed redshift is cosmological in origin. If the redshift distance modulus is used to determine the absolute magnitude of ESO 445-27 for calibrating the TFR, then the absolute magnitude ($M_B$) of this galaxy would be –22.70 compared with a value of –21.53 for the most luminous ScI group calibrator (Russell 2004,2005) and –21.40 predicted from the calibrators. *It should be readily apparent that such contributions to absolute magnitude error introduced by intrinsic redshifts will drive up the observed TFR scatter.*

Federspiel et al (1994) claimed that the wedge pattern of the redshift-distance plot rules out peculiar motions as an explanation and therefore intrinsic TFR scatter is responsible for the large observed scatter. It should be noted that if intrinsic redshifts are the actual cause, then the implication is that intrinsic redshifts increase in size with greater distance.

In fact the observed wedge pattern of the redshift-distance plot is expected in the intrinsic redshift model of Narlikar&Arp (1993 –hereafter NA93) – which predicts the size of a galaxy's intrinsic redshift component is related to the age of the galaxy. NA93 derive a Hubble Constant of ~50 km $s^{-1}$ $Mpc^{-1}$ from the age of the Milky Way. All galaxies the same age as the Milky Way are predicted to follow a tight Hubble relation where $H_0$=50. Galaxies younger than the Milky Way are predicted to have larger observed redshifts than expected for $H_0$=50. The size of this excess is related to each galaxy's age such that the younger the galaxy the larger the excess (intrinsic) redshift component.

For example, ESO 445-27 gives $H_0$=131 km $s^{-1}$ $Mpc^{-1}$. This large $H_0$ value implies ESO 445-27 is younger than the Milky Way in the NA93 model. With a true value of $H_0$=50, the cosmological component of the ESO 445-27 redshift is 4475 km $s^{-1}$ and therefore the remaining intrinsic component is +7247 km $s^{-1}$. A second galaxy the same age as ESO 445-27, but at half the distance (45 Mpc) will be predicted to have an observed redshift that gives an observed $H_0$=131 km $s^{-1}$ $Mpc^{-1}$. However, at this distance the galaxy will have a cosmological redshift component of 2250 km $s^{-1}$ and an intrinsic redshift component of +3646 km $s^{-1}$. Note that while the age of the two galaxies would be the same, the intrinsic component is smaller for the galaxy at the closer distance.

Thus in the NA93 model, the spread in observed redshifts for a sample of galaxies at the same distance results from the absolute age range of the galaxies in the sample rather than superposition of peculiar motions upon expansion of the universe. Making the reasonable assumption that the galaxies over the distance range of the ScI sample plotted in figure 1 have a constant range of ages – and therefore the same range of individual observed $H_0$ values - the size of the intrinsic redshift component is expected to increase with increasing distance.

Figure 3 illustrates how this effect causes a wedge shaped Hubble diagram. Plotted are the 30 galaxies from the Pisces filament sample of Russell (2004,2005). The cluster has a mean distance of ~55 Mpc and the observed redshifts and TFR distances for the individual galaxies are plotted as the central cluster in figure 3. The clusters at ~27 Mpc and ~110 Mpc represent the same 30 galaxies at half and twice their actual distances respectively. Redshifts were recalculated assuming the same individual $H_0$ value (and thus same absolute age in the NA93 model) at the adjusted distances. The resulting wedge shaped Hubble plot is readily apparent.

The ability to predict a wedge shaped Hubble plot should be considered an important test of any mechanisms proposed to explain intrinsic redshifts.

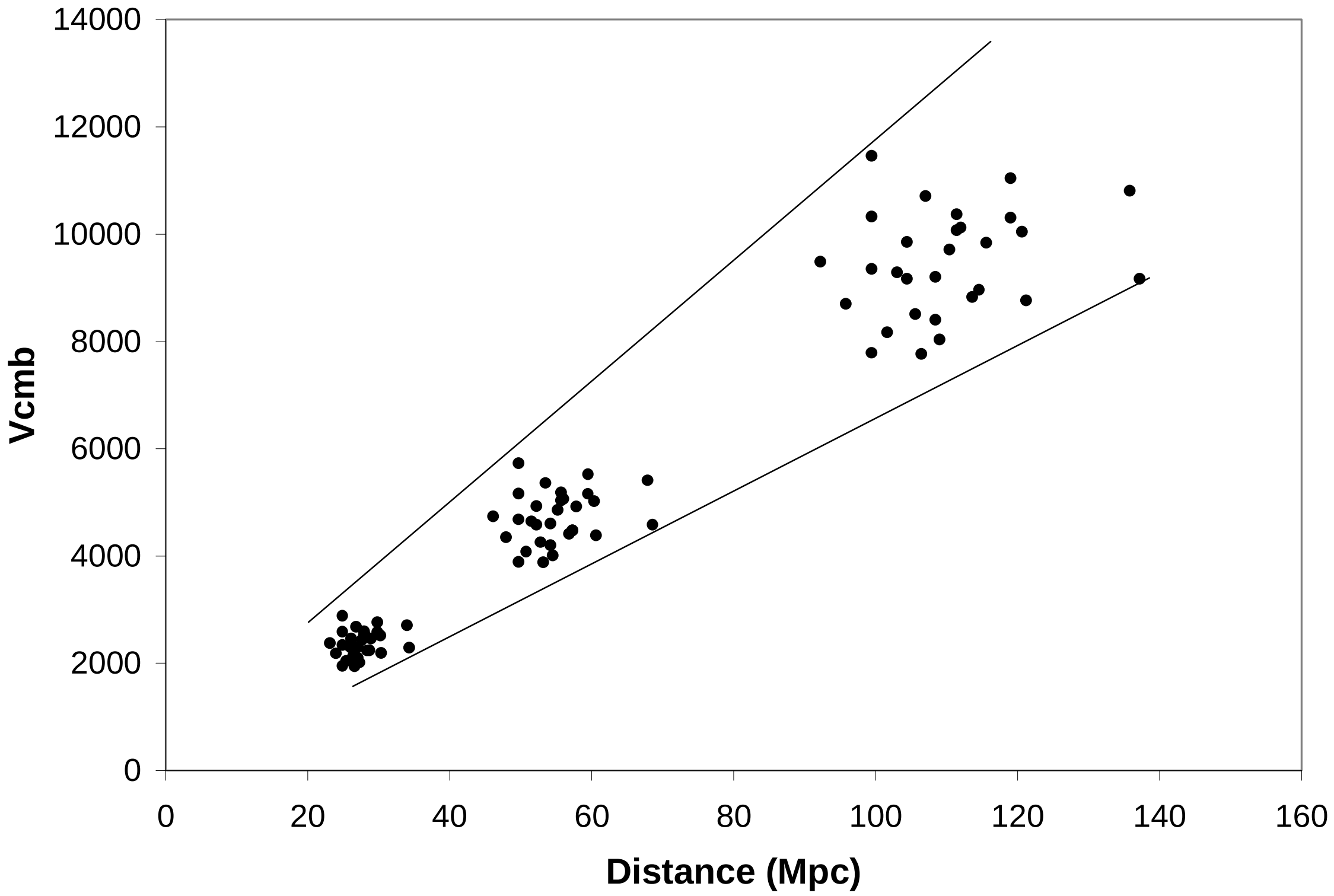

Figure 3 – Redshift distance plot for the Pisces filament sample of Russell (2005) at the TFR distance, and with redshift scaled to half the distance and twice the distance (see text for additional explanation).

## 4. What is the Hubble Constant?

A great difficulty presented by the interpretation that large non-velocity intrinsic redshifts contaminate the measured redshifts is how to determine the correct value of the Hubble Constant – which defines the cosmological component of redshift whether or not an expanding universe model is preferred. The evidence suggests that intrinsic redshifts have a strong tendency to add to the cosmological component of redshift. Thus it is possible that the currently preferred value of $H_0$=72 km $s^{-1}$ $Mpc^{-1}$ (Freedman et al 2001) may be considered an upper limit.

Arp (2002) proposed that the value of $H_0$ could be estimated by using earlier type spiral galaxies (Sab/Sb) which empirically have the least contamination from intrinsic redshifts (Arp 1988, 1990, 1994, 1998) and argued for $H_0$= 55 km $s^{-1}$ $Mpc^{-1}$. It has also been proposed that intrinsic redshifts may be superimposed upon the cosmological redshift component in discrete amounts. From two independent samples (Bell & Comeau 2003; Bell et al 2004) it was found that intrinsic redshifts are superimposed in discrete states upon a Hubble constant of 58 km $s^{-1}$ $Mpc^{-1}$.

What is interesting to note is that the value of $H_0$ found by Arp and Bell is very close to the value advocated by those that argue for large bias corrections to the extragalactic distance scale (eg. Sandage 1999; Theureau et al 1997; Ekholm et al 1999). Recall that these bias corrections are intended to correct for Malmquist bias assuming a large intrinsic scatter in the TFR. In the context of this discussion it can be seen that while these bias corrections are intended to correct for intrinsic luminosity scatter, they may actually be correcting for the effects of intrinsic redshifts. Thus Sandage (1999), Theureau et al (1997), and Ekholm et al (1999) may have derived the correct $H_0$ value from incorrect assumptions.

Finally, it should be noted that Sandage (1994,1999) emphasizes the use of *distance limited* samples in calculating the Hubble Constant, but defines distance limited by setting *redshift limits*. Such redshift limited samples are only true distance limited samples if large non-cosmological intrinsic redshifts do not exist. The result of defining "distance limited" this way is that galaxies which are part of the true distance limited sample but have relatively large intrinsic redshifts will be excluded from the sample and thus a lower value of the Hubble Constant is obtained.

## 5.Conclusion

The observed TFR scatter derived from Cepheid calibrators and cluster samples is significantly smaller than the observed TFR scatter derived from Hubble distances of field galaxies. Three possible explanations for this discrepancy were discussed: (1) field galaxies have significantly larger intrinsic TFR scatter than the calibrators and cluster samples; (2) field galaxies have a velocity dispersion in excess of 1000 km $s^{-1}$; (3) galaxies contain a component of non-velocity intrinsic redshift superimposed upon the cosmological component of redshift. The first possibility has no previous observational basis and is contradicted by the small TFR scatter observed in clusters and calibrators from vastly different environments. The second possibility also has no previous observational basis and is contradicted by the wedge shaped distribution on redshift-distance plots and the small velocity dispersions found in local samples. The third possibility in fact predicts that TFR scatter should be larger when Hubble distances are used to calibrate the TFR. One possible model to explain these results is the existence of an intrinsic redshift component superimposed – possibly in discrete amounts - upon a cosmological component with a Hubble constant of 55 to 60 km $s^{-1}$ $Mpc^{-1}$.

Acknowledgements: This research has made use of the HyperLeda database (Paturel et al 2003) compiled by the LEDA team at the CRAL-Observatoire de Lyon (France). I would like to thank the referee for helpful suggestions that led to improvements in this paper.